\pdfoutput=1

\documentclass[11pt]{article}

\usepackage[final]{acl}

\usepackage{times}
\usepackage{latexsym}

\usepackage[T1]{fontenc}

\usepackage[utf8]{inputenc}

\usepackage{microtype}

\usepackage{inconsolata}

\usepackage{graphicx}
\usepackage{booktabs}
\usepackage{multirow}
\usepackage{graphicx}
\usepackage{makecell}
\usepackage{xcolor}
\usepackage{amsmath}

\title{Who Can Withstand Chat-Audio Attacks? \\ An Evaluation Benchmark for Large Audio-Language Models}

\author{
 \bfseries
 Wanqi Yang$^{1 \ast}$,
 \
 Yanda Li$^{1}$\thanks{Equal contributions},
 \
 Meng Fang$^{2}$,
 \
 Yunchao Wei$^{3}$,
 \
 Ling Chen$^1$
 \\
 \normalsize 
 $ ^1$ University of Technology Sydney,
 $ ^2 $ University of Liverpool,
 $ ^3 $ Beijing Jiaotong University\\
 {\normalsize \tt   wanqi.yang-1@student.uts.edu.au, 
 \normalsize \tt Yanda.Li@student.uts.edu.au}\\
 \normalsize \tt  Meng.Fang@liverpool.ac.uk,
 \normalsize \tt  wychao1987@gmail.com,
 \normalsize \tt ling.chen@uts.edu.au
 }

\begin{document}
\maketitle



\begin{abstract}
Adversarial audio attacks pose a significant threat to the growing use of large audio-language models (LALMs) in voice-based human-machine interactions. While existing research focused on model-specific adversarial methods, real-world applications demand a more generalizable and universal approach to audio adversarial attacks. In this paper, we introduce the Chat-Audio Attacks (CAA) benchmark including four distinct types of audio attacks, which aims to explore the vulnerabilities of LALMs to these audio attacks in conversational scenarios. To evaluate the robustness of LALMs, we propose three evaluation strategies: Standard Evaluation, utilizing traditional metrics to quantify model performance under attacks; GPT-4o-Based Evaluation, which simulates real-world conversational complexities; and Human Evaluation, offering insights into user perception and trust. We evaluate six state-of-the-art LALMs with voice interaction capabilities, including Gemini-1.5-Pro, GPT-4o, and others, using three distinct evaluation methods on the CAA benchmark. Our comprehensive analysis reveals the impact of four types of audio attacks on the performance of these models, demonstrating that GPT-4o exhibits the highest level of resilience. Our data can be accessed via the following link: \href{https://github.com/crystraldo/CAA}{CAA}.

\end{abstract}

\section{Introduction}

Large language models (LLMs) capable of processing text, images, and audio have become increasingly essential for applications that require advanced comprehension and response generation, including customer service~\cite{kolasani2023optimizing,hadi2024large}, automated content creation~\cite{todd2023level,sudhakaran2024mariogpt}, and conversational systems~\cite{he2023large,kopf2024openassistant,yang2025mtpchat}. However, the versatile capabilities of these models also increase their vulnerability to adversarial attacks~\cite{shayegani2023survey,zhao2024evaluating}. This is particularly true in the domain of LLM-driven human-machine voice interaction, where the emergence of such services has accelerated research into audio-based adversarial attacks and defense mechanisms.

Attacks on large audio-language models (LALMs) can cause the models to produce unintended outputs. However, this area has received limited attention, primarily due to the challenges associated with audio as an input modality. Unlike images, audio lacks direct gradient signals, making the crafting of adversarial examples more complex.
Previous research on adversarial audio attacks has focused primarily on targeted attacks~\cite{gong2017crafting,kassis2021practical,zhang2022waveform}, where carefully crafted perturbations are embedded within speech signals. While these samples are effective in misleading models, they often appear as random noise and are easily detectable by human listeners. A notable advancement~\cite{carlini2018audio} introduced a gradient-based optimization approach that utilizes the Connectionist Temporal Classification loss~\cite{graves2006connectionist}—a method designed for time series data in classification tasks. However, this method remains model-specific and lacks broader generalizability.
Universal adversarial audio attacks~\cite{xie2021enabling} are highly relevant to real-world attack scenarios, such as when a speaker makes a verbal error or when they are speaking in a noisy environment. Attackers can pre-design and generate these universal attacks in advance, then apply them to any input audio. Despite their relevance, there has been insufficient exploration of their impact on LALMs.

As LALMs become more prevalent in human-machine voice interactions, the threat posed by these attacks grows significantly. To explore the vulnerabilities of LALMs to adversarial audio attacks, we propose a benchmark of universal adversarial audio attacks specifically based on conversational scenarios, named Chat-Audio Attacks (CAA). The CAA benchmark consists of 360 adversarial audio attack sets, with each set encompassing four distinct types of audio attacks: content attack, emotional attack, explicit noise attack, and implicit noise attack. This results in a total of 1,680 adversarial audio samples.
We believe that CAA benchmark will not only enable researchers to pinpoint weaknesses in LALMs under adversarial audio conditions but also drive the advancement of robust defense mechanisms for LALMs.

\begin{figure*}[ht]
\centering
  \includegraphics[width=1.0\textwidth]{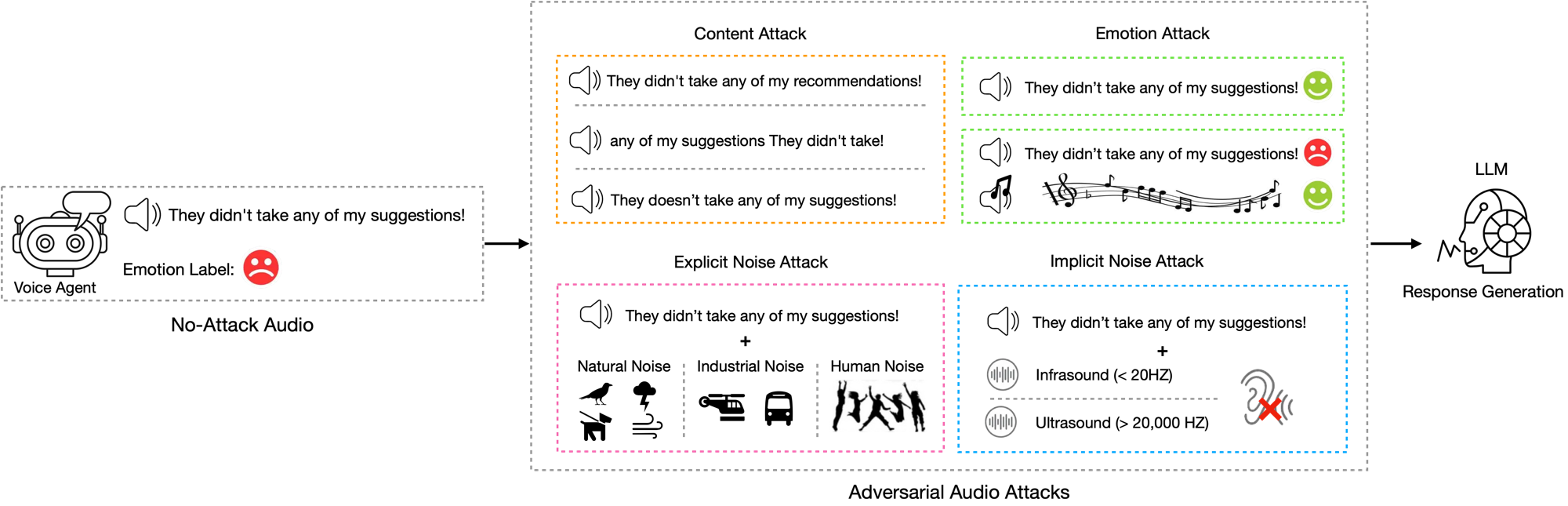}

  \caption{An overview of Chat-Audio Attacks (CAA) benchmark including four distinct types of audio attacks.}
    \label{overview}

 \end{figure*}

In addition, we introduce three evaluation methods to comprehensively assess the resilience of LALMs against adversarial audio attacks: Standard Evaluation, GPT-4o-Based Evaluation, and Human Evaluation. The Standard Evaluation uses rigorous metrics to quantify the accuracy, similarity, and consistency of voice responses under adversarial conditions, providing a repeatable and controlled result. In contrast, the GPT-4o-Based Evaluation simulates real-world interactions, capturing complex, sensitive inaccuracies that standard metrics might overlook. Human Evaluation reflects actual user experience and perceptions, offering crucial insights into user trust.

Finally, we evaluate six state-of-the-art LALMs supporting voice-based conversations on the CAA benchmark, such as Gemini-1.5-Pro~\cite{reid2024gemini} and GPT-4o~\cite{openai_chatgpt}, providing results across the three aforementioned evaluation methods. We analyze the impact of four types of audio attacks on the LALMs and discusse the flaws these models exhibit in the face of such attacks.

The main contributions of this work are summarised as:
\begin{itemize}

    \item We propose a benchmark for universal adversarial audio attacks based on conversation task, called Chat-Audio Attacks (CAA).

    \item We propose three evaluation methods to systematically evaluate the performance of LALMs against adversarial audio attacks.

    \item We perform a comprehensive evaluation of six state-of-the-art LALMs using the CAA benchmark. Based on the three experimental results, we provide an in-depth analysis and discussion of the results.

\end{itemize}
\section{CAA Benchmark}

In CAA benchmark, we target the response generation task by collecting suitable audio for human-machine chat. The overview of CAA benchmark is shown in Figure~\ref{overview}. Each set of audio attack data consists of a quadruplet ($a_i$, $t_i$, ${a_i}^{no\_attack}$ $\mathcal{A}_i$), where $a_i$ represents the original, unprocessed audio containing a single utterance; $t_i$ refers to the transcript of the original audio along with other associated textual labels; ${a_i}^{no\_attack}$ indicates the audio generated by a voice agent reading the transcript without any attack; and the set $\mathcal{A}_i$ includes 3 or 5 types of attack variations of the audio. 

\subsection{Audio Collection}

For unprocessed audio $a_i$ and corresponding transcripts $t_i$ in CAA benchmark, we manually collected data from three publicly available multimodal datasets (text, audio, and visual): MELD, TVQA, and Common Voice.
\begin{itemize}
    
    \item MELD (Multimodal EmotionLines Dataset)~\cite{poria2018meld}: is designed for emotion recognition and classification, derived from the popular TV show Friends. MELD contains numerous dialogue examples, each associated with audio, video, transcripts, and emotion labels (e.g., happiness, sadness, anger, etc.).
    
    \item TVQA~\cite{lei2018tvqa}: primarily focused on understanding video content and associated dialogues in television shows, this dataset covers six famous English-language TV series. Each dialogue instance includes audio, video frames, and transcripts.
    
    \item Common Voice~\cite{ardila2019common}: is a multilingual dataset for speech recognition, provides audios and transcripts. However, the audio samples are not explicitly designed in a dialogue format.
\end{itemize}

After manually filtering and applying GPT-4~\cite{openai_chatgpt} refinement, we collected 120 English speech samples along with their transcriptions from each dataset mentioned above. Notably, the emotional tags from the MELD dataset were also collected to facilitate the generation of emotional attacks in subsequent experiments.

\subsection{Audio Attack Generation}

We processed the collected audio samples to generate five distinct types of audio variations: no attack, content attack, emotional attack, explicit noise attack, and implicit noise attack.

\noindent\textbf{No-Attack Audio} refers to audio generated by a voice agent reading the transcript without any modifications or interference. In CAA benchmark, we utilized AzureSpeechSDK agent~\cite{AzureSpeechSDK} to produce audio recordings. Specifically, for samples sourced from MELD, which include emotion labels, AzureSpeechSDK agent was configured to match the emotional tone indicated by the labels. For TVQA and Common Voice samples, where emotion labels are absent, the agent was instructed to adopt a neutral tone.

We observed that some samples from MELD, TVQA, and Common Voice are often impacted by factors such as speech rate, accent, and clarity, which can obscure the audio information, making them unsuitable as baselines for subsequent comparison and analysis. To address this, we generated no-attack audio to ensure that the LALMs receive clear speech inputs. This serves as a baseline, offering audio free from interference or alterations.

\noindent\textbf{Content Attack} alters a small fraction of the audio's transcribed tokens while preserving the overall semantic meaning. Inspired by these studies~\cite{ribeiro2018semantically,wei2019eda,jin2020bert,li2020bert}, we modified the transcriptions using one of the following strategies: (1) synonym substitution, (2) token rearrangement, or (3) minimal token variation. For synonym substitution, we employed GPT-4 to identify key tokens and replace them with synonyms. For example, ``They didn't take any of my suggestions'' was altered to ``They didn't take any of my recommendations!''. Minimal token variation involved altering non-essential tokens, such as ``didn't'' to ``doesn't''. The modified text was then read aloud by the AzureSpeechSDK agent, preserving the original emotional tone, resulting in content-attacked audio.

The goal of content attacks is to explore whether LALMs are sensitive to token changes or minor errors when the overall meaning of the audio remains preserved.

\noindent\textbf{Emotional Attack} alters the emotional tone of the audio without changing the content. CAA benchmark contains two types of emotional attacks: (1) opposing emotional tone, and (2) opposing emotional background music. In the first scenario, the AzureSpeechSDK agent was instructed to re-read the transcript with an emotion opposite to the original. For instance, if the original sample had an ``angry'' emotion label, the agent would re-read the transcripts with a ``happy'' tone, generating an opposite-emotion audio sample. In the second scenario, we overlaid background music with an opposing emotion onto the no-attack audio and adjusted the music volume to ensure the speaker's voice remained clear. It is important to emphasize that only samples from the MELD dataset in our collection are labeled with emotions. As a result, we utilized 120 samples from MELD to generate two emotional attack audio samples for each, resulting in two distinct emotional tones per sample.

The objective of emotional attacks is to investigate the sensitivity of LALMs to variations in speech emotion and whether a mismatch between speech content and emotional tone influences responses.

\noindent\textbf{Explicit Noise Attack} considers three categories of explicit noise: (1) natural noise (e.g., bird calls, wind, thunder), (2) industrial noise (e.g., car horns, machinery, object collisions), and (3) human noise (e.g., crowd chatter, shouting, laughter). Each noise sample was overlaid on the no-attack audio, with the noise volume adjusted to ensure that the speaker's voice remained clear. We generated 120 samples for each category of explicit noise attack.

Explicit noise attacks are used to evaluate the ability of LALMs to differentiate between the speaker's voice and background noise, as well as to assess their robustness to such interference.

\noindent\textbf{Implicit Noise Attack} indicates human hearing typically ranges from 20 Hz to 20,000 Hz (20 kHz). Sounds outside this range are classified as (1) infrasound, with frequencies below 20 Hz, and (2) ultrasound, with frequencies above 20,000 Hz. We employed the numpy and scipy libraries for digital signal processing, generating infrasound samples at 15 Hz and ultrasonic samples at 22,000 Hz, which were then overlaid onto the no-attack audio. 180 samples were produced for each type of implicit noise attack. It is worth noting that we deliberately increased the volume of the implicit noise. However, since these sound waves fall outside the normal auditory range of human hearing, their addition to the mixed audio did not compromise the clarity of the speaker's voice.

The objective of implicit noise attacks is to assess whether LALMs, similar to humans, remain unaffected by inaudible noise.

\subsection{Quality Control}

In the \textbf{data collection} phase, we identified several unqualified samples from the MELD, TVQA, and Common Voice datasets. These included: 1) non-English; 2) containing sensitive topics; 3) reasonable responses could not be generated. To address this, we established the following criteria for manual sample collection: 1) the speech must be in English; 2) it must not contain sensitive topics such as sex, drugs, or religion; 3) it must have a minimum of six words; 4) it should not consist of simple greetings or farewells; 5) it should not reference unfamiliar names, places, or institutions; 6) it should avoid professional terminology; and 7) no pronouns like "this" should be used.

To further ensure the respondability of the audio content, we employed GPT-4 for an additional filtering step. The speech transcript was input into GPT-4, and responses were generated based on the designed prompt, as shown in Appendix~\ref{sec:appendixA}. If GPT-4 failed to provide a reasonable response, the sample was discarded. Ultimately, we collected a total of 360 high-quality English audio samples along with their corresponding transcriptions.

Moreover, in the \textbf{data generation} phase, we placed significant emphasis on the quality of the generated audio. Initially, we observed that some samples from the MELD, TVQA, and Common Voice datasets were frequently affected by factors such as speech rate, accent, and clarity, obscuring important audio information. To address this, we utilized AzureSpeechSDK agent to re-synthesize the audio, adjusting the speech rate to be slower and increasing the volume for better clarity. The quality of the no-attack audio was manually verified to ensure it met high standards. These high-quality no-attack samples not only serve as a baseline but also provide a solid foundation for generating attack samples. Furthermore, we adjusted the volume of background music and noise to ensure that the human voice remained clearly audible to listeners.

\subsection{Benchmark Statistics}

The CAA benchmark comprises 360 sets of audio attack data ($a_i$, $t_i$, ${a_i}^{no_attack}$ $\mathcal{A}_i$), resulting in a total of 1,680 samples across five distinct types of audio attacks. On average, each audio sample contains 10 tokens. Our benchmark encompasses six emotional labels: surprise, sadness, joy, anger, fear, and disgust. Additionally, we provide generation scripts for the five types of audio attacks, encouraging researchers to produce more samples for evaluation. The Table~\ref{benchmark} below summarizes the number of samples for each audio attacks in the CAA benchmark.

\vspace{-2mm}
\begin{table}[h]

\centering
\resizebox{\linewidth}{!}{
\begin{tabular}{l|c|m{1.5cm}<{\centering}|m{1.5cm}<{\centering}|m{1.5cm}<{\centering}}
\toprule
\multicolumn{2}{c|}{\textbf{Audio Attack}} &\textbf{MELD} &\textbf{TVQA}& \textbf{Common Voice}\\ 
\midrule
\multicolumn{2}{c|}{No Attack}&120 &120& 120\\
\midrule
\multicolumn{2}{c|}{Content Attack}&120 &120& 120\\
\midrule
\multirow{2}*{Emotion Attack} &{Opp-Emo Tone} & 120& - & - \\
\cline{2-5}
\multirow{2}*{ } &{Opp-Emo Music} & 120& - & - \\
\midrule
\multirow{3}*{Explicit Noise} &{Natural Noise}&40&40&40 \\
\cline{2-5}
\multirow{3}*{ } &{Industrial Noise}&40&40&40 \\
\cline{2-5}
\multirow{3}*{ } &{Human Noise}&40&40&40 \\
\midrule
\multirow{2}*{Implicit Noise} &{Infrasound } & 60& 60 & 60 \\
\cline{2-5}
\multirow{2}*{ } &{Ultrasound} & 60& 60 & 60 \\
\midrule
\multicolumn{2}{c|}{Total}&\multicolumn{3}{c}{1,680}\\

\bottomrule
\end{tabular} 
}
\vspace{-2mm}
\caption{\label{benchmark}
CAA benchmark statistics including five distinct types of audio attacks.
}
\vspace{-3mm}
\end{table}
\section{Experiments}
\subsection{Experimental Setup}

\textbf{Models} We present a comprehensive performance evaluation of the most popular large audio-language models, including SpeechGPT~\cite{zhang2023speechgpt}, SALMONN~\cite{tang2023salmonn}, Qwen2-Audio~\cite{chu2024qwen2}, LLama-Omni~\cite{fang2024llama}, Gemini-1.5-pro~\cite{reid2024gemini} and GPT-4o~\cite{achiam2023gpt}.
\begin{table*}[t]

\centering
\resizebox{\linewidth}{!}{
\begin{tabular}{l|c|c|c|c|c|c|c|c|c}
\toprule
\multirow{2}*{\textbf{Model}} & \multirow{2}*{\textbf{Metrics}}& \multirow{2}*{\textbf{Content Attack}} &\multicolumn{2}{c|}{\textbf{Emotion Attack}} & \multicolumn{3}{c|}{\textbf{Explicit Noise}}& \multicolumn{2}{c}{\textbf{Implicit Noise}}\\ 
\cline{4-10}
\multirow{2}{*}{}&\multirow{2}{*}{}&\multirow{2}{*}{} &\textbf{Opp-Emo Tone}&\textbf{Opp-Emo Music}&\textbf{Natural Noise}&\textbf{Industrial Noise}&\textbf{Human Noise}&\textbf{Infrasound}&\textbf{Ultrasound}\\
\midrule
\multirow{3}*{SpeechGPT} &{WER (↓)} & 1.79 & 1.76 & 1.74 & 2.25 & 1.94 & 1.29 & 2.21 &  1.28\\
\cline{2-10}
\multirow{3}*{} &{ROUGE-L (↑)} & 0.17 & 0.18 & 0.12 & 0.12 & 0.10 & 0.10 & 0.14 & 0.20 \\
\cline{2-10}
\multirow{3}*{} &{COS (↑)} & 0.23 & 0.24 & 0.15 & 0.16 & 0.13 & 0.14 & 0.19 & 0.26 \\
\midrule[1.5pt]
\multirow{3}*{SALMONN} &{WER (↓)} & \textbf{0.80} & 1.31 & 1.00 & 0.65 & 1.06 & 1.19 & 1.46 & 0.61\\
\cline{2-10}
\multirow{3}*{} &{ROUGE-L (↑)} & \textbf{0.63} & \textbf{0.61} & 0.54 & \textbf{0.68} & \textbf{0.57} & \textbf{0.57} & \textbf{0.58} & 0.74 \\
\cline{2-10}
\multirow{3}*{} &{COS (↑)} & \textbf{0.69} & \textbf{0.69} & 0.60 & \textbf{0.75} & \textbf{0.65} & \textbf{0.63} & \textbf{0.65} & 0.78 \\
\midrule[1.5pt]
\multirow{3}*{Qwen2-Audio} &{WER (↓)} & 1.59 & 1.27 & 1.24 & 1.92 & 1.21 & 1.10 & 1.80 & 0.95 \\
\cline{2-10}
\multirow{3}*{} &{ROUGE-L (↑)} & 0.38 & 0.32 & 0.38 & 0.36 & 0.36 & 0.36 & 0.36 & 0.54 \\
\cline{2-10}
\multirow{3}*{} &{COS (↑)} & 0.52 & 0.45 & 0.51 & 0.51 & 0.50 & 0.50 & 0.49 & 0.65 \\
\midrule[1.5pt]
\multirow{3}*{LLama-Omni} &{WER (↓)} & 1.04 & \textbf{0.91} & \textbf{0.65} & \textbf{0.64} & \textbf{0.77} & \textbf{0.96} & \textbf{0.67} & \textbf{0.37} \\
\cline{2-10}
\multirow{3}*{} &{ROUGE-L (↑)} & 0.36 & 0.38 & \textbf{0.56} & 0.58 & \textbf{0.57} & 0.42 & 0.56 &  \textbf{0.75}\\
\cline{2-10}
\multirow{3}*{} &{COS (↑)} & 0.45 & 0.46 & \textbf{0.63} & 0.64 & 0.62 & 0.51 & 0.63 & \textbf{0.79} \\
\midrule[1.5pt]
\multirow{3}*{Gemini-1.5-Pro} &{WER (↓)} & 1.34 & 1.20 & 1.27 & 1.34 & 1.36 & 1.50 & 1.31 & 1.31 \\
\cline{2-10}
\multirow{3}*{} &{ROUGE-L (↑)} & 0.17 & 0.17 & 0.24 & 0.21 & 0.24 & 0.21 & 0.21 & 0.22 \\
\cline{2-10}
\multirow{3}*{} &{COS (↑)} & 0.25 & 0.25 & 0.32 & 0.30 & 0.35 & 0.27 & 0.30 & 0.31 \\
\midrule[1.5pt]
\multirow{3}*{GPT-4o} &{WER (↓)} & 1.12  & 1.07 & 1.10 & 1.18 & 1.36 & 1.11 & 1.25 & 1.13 \\
\cline{2-10}
\multirow{3}*{} &{ROUGE-L (↑)} & 0.25 & 0.25 & 0.25 & 0.20 & 0.17 & 0.23 & 0.22 & 0.17 \\
\cline{2-10}
\multirow{3}*{} &{COS (↑)} & 0.39 & 0.39 & 0.39 & 0.33 & 0.27 & 0.37 & 0.35 & 0.28 \\

\bottomrule
\end{tabular} 
}

\caption{\label{Standard Evaluation}
Standard evaluation results on CAA benchmark. Performance comparison of the LALMs under various adversarial conditions using \textit{WER}, \textit{ROUGE-L}, and \textit{COS} metrics. 
}

\end{table*}
\textbf{Inference Setup} For model inference, we adopt a zero-shot setup, where the CAA samples are directly fed into the models. SpeechGPT and Qwen2-Audio natively support chat functionality, allowing direct input of audio for generating response. For SALMONN and LLama-Omni, we format questions according to their ``Model Prompts Guide'' to facilitate the Q\&A process. The inference for these models are conducted on a single A100-80G GPU. For GPT-4o and Gemini, we utilize their API interfaces, setting up specific prompts to conduct the inference.

\textbf{Evaluation Methods} The evaluation is conducted from three key perspectives: standard evaluation, GPT4o-based evaluation, and human evaluation. In these evaluation methods, all audio content is presented in the form of transcribed text. 

We collect all prediction results and evaluate them based on the three aforementioned evaluation methods. Detailed configurations for the models and prompts are provided in the Appendix~\ref{sec:appendixB}.

\begin{table*}[t]

\centering
\resizebox{\linewidth}{!}{
\begin{tabular}{l|c|c|c|c|c|c|c|c|c}
\toprule
\multirow{2}*{\textbf{Model}} & \multirow{2}*{\textbf{Metrics}}& \multirow{2}*{\textbf{Content Attack}} &\multicolumn{2}{c|}{\textbf{Emotion Attack}} & \multicolumn{3}{c|}{\textbf{Explicit Noise}}& \multicolumn{2}{c}{\textbf{Implicit Noise}}\\ 
\cline{4-10}
\multirow{2}{*}{}&\multirow{2}{*}{}&\multirow{2}{*}{} &\textbf{Opp-Emo Tone}&\textbf{Opp-Emo Music}&\textbf{Natural Noise}&\textbf{Industrial Noise}&\textbf{Human Noise}&\textbf{Infrasound}&\textbf{Ultrasound}\\
\midrule
\multirow{4}*{SpeechGPT} &{NC (↑)} &  \multicolumn{8}{c}{2.39} \\
\cline{2-10}
\multirow{4}*{} &{ACoh (↑)} & 1.76 & 1.49 & 1.40 & 1.32 & 1.24 & 1.24 & 1.23 & 1.86 \\
\cline{2-10}
\multirow{4}*{} &{ACor (↑)} & 1.58 & 1.39 & 1.37 & 1.23 & 1.18 & 1.16 & 1.15 & 1.67 \\
\cline{2-10}
\multirow{4}*{} &{LR (↑)} & 2.65 & 2.15 & 2.08 & 2.10 & 1.89 & 2.12 & 2.13 & 2.71 \\
\midrule[1.5pt]
\multirow{4}*{SALMONN} &{NC (↑)} &  \multicolumn{8}{c}{2.13} \\
\cline{2-10}
\multirow{4}*{} &{ACoh (↑)} & 1.98 & 2.14 & 1.93 & 1.48 & 1.64& 1.84 & 1.56 & 1.82 \\
\cline{2-10}
\multirow{4}*{} &{ACor (↑)} & 2.01 & 2.20 & 1.97 & 1.60 & 1.80 & 1.86 & 1.55 & 2.11 \\
\cline{2-10}
\multirow{4}*{} &{LR (↑)} & 2.78 & 3.08 & 3.15 & 2.26 & 2.68 & 2.88 & 2.37 & 2.84 \\
\midrule[1.5pt]
\multirow{4}*{Qwen2-Audio} &{NC (↑)} &  \multicolumn{8}{c}{3.46} \\
\cline{2-10}
\multirow{4}*{} &{ACoh (↑)} & 2.90 & 2.71 & 2.85 & 2.31 & 2.5 & 2.78 & 2.41 & 3.04 \\
\cline{2-10}
\multirow{4}*{} &{ACor (↑)} & 2.53 & 2.36 & 2.64 & 1.99 & 2.28 & 2.48 & 2.15 & 2.92 \\
\cline{2-10}
\multirow{4}*{} &{LR (↑)} & 4.02 & 4.05 & 4.13 & 3.24 & 3.80 & 4.08 & 3.49 & 4.06 \\
\midrule[1.5pt]
\multirow{4}*{LLama-Omni} &{NC (↑)} &  \multicolumn{8}{c}{3.50} \\
\cline{2-10}
\multirow{4}*{} &{ACoh (↑)} & 3.05 & 3.08 & 3.24 & \textbf{2.72} & 3.11 & 2.95 & 2.79 & \textbf{3.31} \\
\cline{2-10}
\multirow{4}*{} &{ACor (↑)} & 2.62 & 2.76 & 3.14 & \textbf{2.62} & \textbf{3.02} & 2.68 & \textbf{2.66} & \textbf{3.53} \\
\cline{2-10}
\multirow{4}*{} &{LR (↑)} & 4.31 & 4.32 & 4.37 & \textbf{3.59} & 4.26 & 4.33 & 3.80 & 4.34 \\
\midrule[1.5pt]
\multirow{4}*{Gemini-1.5-Pro} &{NC (↑)} &  \multicolumn{8}{c}{3.58} \\
\cline{2-10}
\multirow{4}*{} &{ACoh (↑)} & 3.15 & 3.30 & 3.32 & 2.42 & \textbf{3.21} & 3.10 & 2.78 & 2.95 \\
\cline{2-10}
\multirow{4}*{} &{ACor (↑)} & 2.62 & 2.72 & 2.69 & 2.00 & 2.78 & 2.75 & 2.28 & 2.59 \\
\cline{2-10}
\multirow{4}*{} &{LR (↑)} & 4.21 & 4.13 & 4.26 & 3.10 & 4.24 & 4.22 & 3.55 & 4.00 \\
\midrule[1.5pt]
\multirow{4}*{GPT-4o} &{NC (↑)} &  \multicolumn{8}{c}{\textbf{4.45}} \\
\cline{2-10}
\multirow{4}*{} &{ACoh (↑)} & \textbf{3.94} & \textbf{4.35} & \textbf{4.43} & 2.57 & 3.01 & \textbf{3.37} & \textbf{3.02} &  2.70 \\
\cline{2-10}
\multirow{4}*{} &{ACor (↑)} & \textbf{3.36} & \textbf{3.56} & \textbf{3.61} & 2.15 & 2.52 & \textbf{2.80} & 2.49 & 2.26 \\
\cline{2-10}
\multirow{4}*{} &{LR (↑)} & \textbf{4.80} & \textbf{4.80} & \textbf{4.82} & 3.41 & \textbf{4.78} & \textbf{4.85} & \textbf{3.89} & \textbf{4.78} \\

\bottomrule
\end{tabular} 
}

\caption{\label{GPT-4o-Based Evaluation}GPT-4o-based evaluation results on CAA benchmark. Performance comparison of the LALMs under various adversarial conditions using NC, ACoh, ACor and LR metrics.
}

\end{table*}

\subsection{Standard Evaluation}

In this section, we evaluate the models by comparing their outputs on responses to no-attack audio with attacked audio using three key metrics: \textit{WER}, \textit{ROUGE-L}~\cite{lin2004rouge}, and \textit{COS} (Cosine Similarity). This rigorous, metric-based approach quantifies the similarity and consistency of the models' responses under various adversarial conditions, providing a controlled and repeatable framework for analyzing system performance and measuring robustness.

\begin{itemize} 
    
    \item \textit{WER}: WER measures the discrepancy between the responses to no-attack audio and attacked audio by quantifying the proportion of differing words. A lower WER score indicates that the model generates more consistent responses, even under adversarial conditions.
    
    \item \textit{ROUGE-L}: ROUGE-L assesses the overlap between the two sets of responses, focusing on the longest common subsequences. A higher ROUGE-L score reflects the model's ability to retain essential information and structure when facing adversarial attacks. 
    
    \item \textit{COS}: COS measures the semantic similarity between the output from no-attack audio and attacked audio. A higher COS score indicates that the model maintains semantic consistency even when adversarial noise is introduced.

\end{itemize}

As shown in Table~\ref{Standard Evaluation}, presenting the performance of the models across these metrics and comparing how well they handle adversarial interference on the audio inputs.

\subsection{GPT-4o-Based Evaluation}
The GPT-4o-Based Evaluation utilizes a more sophisticated set of criteria, leveraging the advanced capabilities of GPT-4o to simulate real-world interaction scenarios and evaluate the effects of adversarial attacks on model behavior. This evaluation is designed to capture more complex, context-sensitive inaccuracies that might escape more conventional metric-driven assessments, offering insights into the model's performance in dynamically changing environments.

In this evaluation, we compare model responses to no-attack and attacked audio across four key metrics, each rated on a scale from 1 to 5. Higher scores indicate better performance and greater resilience to adversarial attacks, with detailed prompt settings provided in the Appendix~\ref{sec:appendixA}.

\begin{itemize}

    \item \textit{No-attack Coherence (NC)}: This metric evaluates how well the no-attack response meaningfully and adequately answers the question or prompt posed by the original audio transcript. A higher score (closer to 5) signifies a strong alignment, while a lower score (closer to 1) indicates that the response deviates significantly from the expected meaning. If the NC score is 1, the remaining metrics (ACoh, ACor, and LR) are automatically rated as 1, reflecting an overall failure in response quality.

    \item \textit{Attack Coherence (ACoh)}: This metric assesses how well the attacked response continues to meaningfully and adequately answer the original question or prompt posed by the audio transcript, despite the attack. A higher score suggests that the model continues to generate coherent and contextually relevant responses, while a lower score indicates significant degradation in relevance due to the attack.

    \item \textit{Attack Correlation (ACor)}: This metric measures the correlation between the attacked response and the no-attack response. A higher score indicates that the core meaning of the no-attack response is retained, while a lower score suggests that the attack has caused notable alterations to the response content.

    \item \textit{Linguistic Robustness (LR)}: This assesses whether the attacked response maintains grammatical correctness, sentence continuity, and logical flow. A higher score indicates that the model preserves linguistic structure even under attack, while a lower score reflects disruptions in coherence or grammatical errors.
\end{itemize}
\vspace{-2mm}
 Table~\ref{GPT-4o-Based Evaluation} presents the evaluation results for each model, comparing their performance on no-attack and attacked audio inputs.


\subsection{Human Evaluation}

In addition to automated evaluations, we conducted a human evaluation to assess the models' performance to reflect actual user experience and perception. It is essential for understanding the practical implications of adversarial attacks, particularly in terms of user satisfaction and trust.

The evaluation was carried out by five native English-speaking university students (three male and two female). Each evaluator independently rated the models' outputs using the same \textit{No-attack Coherence (NC)} and \textit{Attacked Coherence (ACoh)} metrics as defined in the GPT-4o-Based Evaluation. Both metrics were scored on a scale from 1 to 5, where higher scores indicate better performance and greater resilience to adversarial conditions. To ensure consistency in scoring, all evaluators followed standardized testing guidelines, and the final scores were averaged across the five evaluators. This human assessment helps ensure the reasonableness and relevance of the automated results.

The evaluations were conducted in a controlled environment, ensuring a consistent testing setup for all evaluators. By averaging the scores across all evaluators, we ensure that the results reflect a balanced and comprehensive assessment of the models’ performance in both no-attack and adversarial conditions.

Table~\ref{Human Evaluation} presents the human evaluation scores for each model, reflecting their performance in both no-attack and adversarial conditions.

\begin{table}[t]

\centering
\resizebox{\linewidth}{!}{
\begin{tabular}{l|m{1.5cm}<{\centering}|m{1.1cm}<{\centering}|m{1.2cm}<{\centering}|m{1.2cm}<{\centering}|m{1.2cm}<{\centering}}
\toprule
{Model} &{Metrics} & {Content Attack} & {Emotion Attack} &{Explicit Noise} & {Implicit Noise} \\ 
\midrule
\multirow{2}*{SpeechGPT} & {NC (↑)} &  \multicolumn{4}{c}{2.52} \\
\cline{2-6}
\multirow{2}*{} & {ACoh (↑)} & 2.12 & 1.78 & 1.43 & 1.66 \\
\midrule[1.5pt]

\multirow{2}*{SALMONN} & {NC (↑)} &  \multicolumn{4}{c}{2.04} \\
\cline{2-6}
\multirow{2}*{} & {ACoh (↑)} & 2.03 & 2.25 & 1.98 & 1.55 \\
\midrule[1.5pt]

\multirow{2}*{Qwen2-Audio} & {NC (↑)} &  \multicolumn{4}{c}{3.82} \\
\cline{2-6}
\multirow{2}*{} & {ACoh (↑)} & 3.02 & 2.88 & 2.34 & 2.77 \\
\midrule[1.5pt]

\multirow{2}*{LLama-Omni} & {NC (↑)} &  \multicolumn{4}{c}{3.75} \\
\cline{2-6}
\multirow{2}*{} & {ACoh (↑)} & 3.40 & 3.22 & 2.88 & 3.15   \\
\midrule[1.5pt]

\multirow{2}*{Gemini-1.5-Pro} & {NC (↑)} &  \multicolumn{4}{c}{3.92} \\
\cline{2-6}
\multirow{2}*{} & {ACoh (↑)} & 3.20 & 3.41 & 3.24 & 2.87 \\
\midrule[1.5pt]

\multirow{2}*{GPT-4o} & {NC (↑)} &  \multicolumn{4}{c}{4.33} \\
\cline{2-6}
\multirow{2}*{} & {ACoh (↑)} & 3.88 & 4.12 & 3.27 & 3.08 \\
\bottomrule
\end{tabular}
}

\caption{\label{Human Evaluation} Human evaluation results on CAA benchmark. Metrics include NC (No-attack Coherence) and ACoh (Attacked Coherence).}

\end{table}

\section{Discussion}

\textbf{Are LALMs sensitive to token changes or minor errors?}

It is evident that different LALMs exhibit varying degrees of sensitivity to token changes or minor errors. GPT-4o consistently shows strong robustness across most metrics (WER, ROUGE-L, COS, ACoh, ACor, and LR), indicating lower sensitivity to token-level adversarial attack. In contrast, SpeechGPT and Qwen2-Audio exhibit greater vulnerability, with lower scores in these key areas, suggesting that minor token changes can significantly degrade their performance. 

\textbf{Is it good news that LALMs are unaffected by the mismatch between speech content and emotional tone?}

We argue that it is \textbf{not} good news that LALMs remain unaffected by emotional mismatches. Although large language models demonstrate resilience by maintaining high levels of coherence, correlation, and semantic similarity, this also reflects their relative weakness in emotional awareness. Current LALMs still have considerable scope for improvement in recognizing emotional subtleties, as humans can easily detect emotional mismatches, such as sarcasm or passive-aggressive tones in conversations. While SpeechGPT is notably impacted by mismatches between speech content and emotional tone, this does not indicate a heightened sensitivity to emotional shifts, as its overall coherence score remains relatively low.

\textbf{Which explicit noise attacks have the most significant impact on LALMs?}

Natural noise has the most significant overall impact on LALMs across all metrics, especially on SpeechGPT and SALMONN, which shows the highest sensitivity to it. Industrial noise also causes notable attacks but is handled better by LALMs like Llama-Omni and Gemini-1.5-Pro. Human noise, while still impactful, is generally less detrimental compared to the other explicit noises. Overall, SpeechGPT and SALMONN show the most vulnerability across all types of explicit noise attacks, while Llama-Omni, Gemini-1.5-Pro and GPT-4o demonstrate stronger robustness.

\textbf{Are LALMs unaffected by inaudible noise?}

None of the models remain entirely unaffected, especially infrasound, which has a greater impact on accuracy (WER), semantic similarity (COS), coherence (ACoh), and grammatical structure (LR). In comparison to ultrasound, infrasound emerges as the more detrimental form of implicit noise, with models like SpeechGPT, SALMONN, Gemini-1.5-Pro and GPT-4o showing significant vulnerability to these attacks. However, Llama-Omni demonstrate greater robustness, performing consistently better across all metrics and handling both types of implicit noise more effectively.

\textbf{What helps models stay robust against adversarial audio?}

The results indicate that all models are affected by adversarial attacks, especially by Explicit Noise and Implicit Noise, which cause a significant number of prediction errors. The evaluation reveals that SpeechGPT and SALMONN demonstrate relatively weak robustness across various adversarial scenarios, exhibiting significant performance degradation when facing different adversarial audio attacks. In contrast, models like Qwen2-Audio, LLama-Omni, and Gemini-1.5-Pro demonstrate stronger resilience, particularly when dealing with emotional attacks and implicit noise. These models manage to maintain logical coherence and linguistic accuracy, with LLama-Omni and Gemini-1.5-Pro standing out for their robust performance across various adversarial conditions.

However, \textbf{GPT-4o clearly emerges as the best-performing model overall}. It consistently delivers coherent, contextually relevant, and linguistically robust responses, even under severe adversarial conditions. The model's ability to handle different types of attacks highlights its superior robustness and adaptability, which can be attributed to its extensive pre-training on large-scale datasets. This factor allow GPT-4o to better understand and process a wide variety of inputs, making it more resistant to adversarial perturbations.

\textbf{How architectural or training differences contribute to robustness?}

1) Vulnerability of Models with Transcription Modules: Models such as SpeechGPT, which incorporate transcription modules, are particularly fragile. These models first transcribe input audio into text before performing inference. Consequently, the accuracy of the transcription process becomes critical. Our adversarial samples significantly degrade the performance of these transcription modules, leading to substantial drops in transcription accuracy when processing audio with noise.

2) Training Limitations of SALMONN: SALMONN lacks a task specifically designed for generating responses during training. As a result, both human and automated evaluations rated SALMONN’s responses poorly in terms of satisfaction. This shortcoming causes the model to behave inconsistently under attack: at times transcribing the audio, other times describing its content, and occasionally speculating about the speaker’s emotions. This inconsistency further undermines its robustness.

3) Robustness of Qwen2-Audio and Llama-Omni: Unlike SpeechGPT and SALMONN, Qwen2-Audio and Llama-Omni do not rely on transcribing input audio into text. Moreover, their training datasets include audio with noise, such as laughter, which contributes to their superior robustness against adversarial attacks.

4) Observations on GPT-4o and Gemini-1.5-Pro: While GPT-4o and Gemini-1.5-Pro have not released their model codes, an intriguing observation is their heightened vulnerability to infrasound and ultrasound attacks. In contrast, SALMONN is minimally affected by such attacks. This highlights an area for improvement in these models.

\textbf{How these findings might influence model design or inform strategies for improving adversarial robustness?}

Based on the analysis of the model structure and training methods presented above, we propose several directions for future research on speech models:

1) Developing enhanced transcription modules that demonstrate greater robustness to audio perturbations, or exploring alternative approaches that eliminate the reliance on transcription altogether.

2) Incorporating multi-task training objectives to improve the adaptability and generalizability of speech models.

3) Diversifying training datasets to better reflect real-world scenarios and adversarial conditions, ensuring broader applicability and resilience.

4) Designing targeted defense mechanisms to counter specific adversarial techniques, such as infrasound and ultrasound attacks. These directions aim to address existing challenges while paving the way for more robust and versatile speech models.




\section{Related works}
\subsection{Audio/Speech Language Models}
In the field of LALMs, initial systems~\cite{lakhotia2021generative, radford2023robust, borsos2022audiolm} utilized either acoustic or semantic tokens to enable generation from audio inputs into text or audio outputs. With the technological advancements brought by LLMs, the recent trend has shifted towards multimodal models~\cite{tang2023salmonn,chen2023x, wu2023decoder,fathullah2024prompting} are leveraging the combined strengths of both speech and text modalities, substantially enhancing the versatility and effectiveness of audio-based applications.

Models like SpeechGPT~\cite{zhang2023speechgpt} utilize a cross-modal architecture that aligns speech and text for tasks such as instruction following and spoken dialogue. SALMONN~\cite{tang2023salmonn} introduces dual encoders to process diverse audio inputs, excelling in speech recognition and even audio storytelling. Qwen2-Audio~\cite{chu2024qwen2}, LLama-Omni~\cite{fang2024llama}, and Gemini-1.5-pro~\cite{reid2024gemini} each contribute unique capabilities ranging from voice chat and low-latency interactions to handling complex multimodal data. Additionally, GPT-4o~\cite{achiam2023gpt} expands upon these functionalities by ensuring robust performance in audio-text interactions within noisy environments, marking a significant milestone in the field.

\subsection{Audio Attacks}

In the domain of adversarial attacks, the concept was first pioneered in the field of image processing~\cite{goodfellow2014explaining}, where slight perturbations to input pixels~\cite{szegedy2013intriguing} could mislead traditional neural network models~\cite{krizhevsky2012imagenet} into producing incorrect results. This methodological foundation laid the groundwork for similar explorations in the audio domain, particularly targeting systems such as automatic speech recognition (ASR)~\cite{carlini2018audio,neekhara2019universal} and spoofing/automatic speaker verification (ASV)~\cite{xie2020real,kassis2021practical,zhang2022waveform}, where security and reliability are critical.

The initial generation of adversarial samples utilized optimization methods first developed for music genre classification~\cite{kereliuk2015deep}. These techniques manipulated entire audio waveforms to avoid detection, altering not only specific acoustic features but the entire sound profile while preserving perceptual quality. In contrast, in the field of speech paralinguistics~\cite{gong2017crafting,kassis2021practical,zhang2022waveform}, the Fast Gradient Sign Method has been employed to craft adversarial samples aimed at disrupting systems. 
LLMs that process diverse data types such as text, images, and audio offer enhanced capabilities for generating human-like responses across various applications. However, their multi-modal nature also increases vulnerability to jailbreaks~\cite{audio_based_jailbreak_2023} and adversarial attacks, with potential exploits spanning across all processed modalities, allowing attackers to bypass safety constraints embedded within these models.

\section{Conclusion}
This work explored the vulnerabilities of LALMs to adversarial audio attacks in conversational scenarios. We introduced the Chat-Audio Attacks (CAA) benchmark, consisting of 360 adversarial attack sets across four attack types: content, emotional, explicit noise, and implicit noise attacks. Our evaluation of six state-of-the-art LALMs using three methods—Standard Evaluation, GPT-4o-Based Evaluation, and Human Evaluation—revealed and discussed significant model vulnerabilities under adversarial conditions.

The CAA benchmark highlights these weaknesses and provides a foundation for developing more robust defense mechanisms. As LALMs are increasingly integrated into voice interactions, enhancing their resilience against adversarial audio attacks remains a crucial area for future research.
\section{Limitations}
Despite the comprehensive design of the Chat-Audio Attacks (CAA) benchmark, our work is not without limitations. First, while we have created a diverse set of adversarial audio samples covering four distinct types of attacks, these scenarios are based on controlled conversational settings and may not capture all the complexities of real-world environments. This could limit the generalizability of our findings in highly dynamic or diverse speech environments.

Meanwhile, there is very limited research on audio jailbreak attacks, and there is almost no open-source work available in this area. As a result, we were unable to generate corresponding adversarial samples for testing such attacks. This represents a significant gap in the exploration of both white-box and black-box attacks on specific LALMs. There is considerable room for future development in addressing the security vulnerabilities posed by these types of attacks, which remain an important but underexplored aspect of LALM security.

\section{Ethics Statement}
This research explores vulnerabilities in LALMs through adversarial audio attacks with the goal of improving model robustness. All adversarial samples were generated solely for research purposes to enhance the security of LALM-based systems. Human evaluation was conducted with informed consent, and no personally identifiable information was collected. We prioritize ethical considerations, ensuring that the work contributes to safer and more reliable conversational AI technologies.

\section{Acknowledgements}
This project is partially supported by ARC DP240101349.

\bibliography{custom}

\clearpage
\appendix
\section{Model Configurations}
\label{sec:appendixB}
Table \ref{table:models_overview} provides an overview of the key information for the evaluated models, including their parameter sizes, language models, audio models, and the prompts used for inference.

\section{Prompt Description}
\label{sec:appendixA}
\begin{figure*}[t]
\centering
\includegraphics[width=1.0\textwidth]{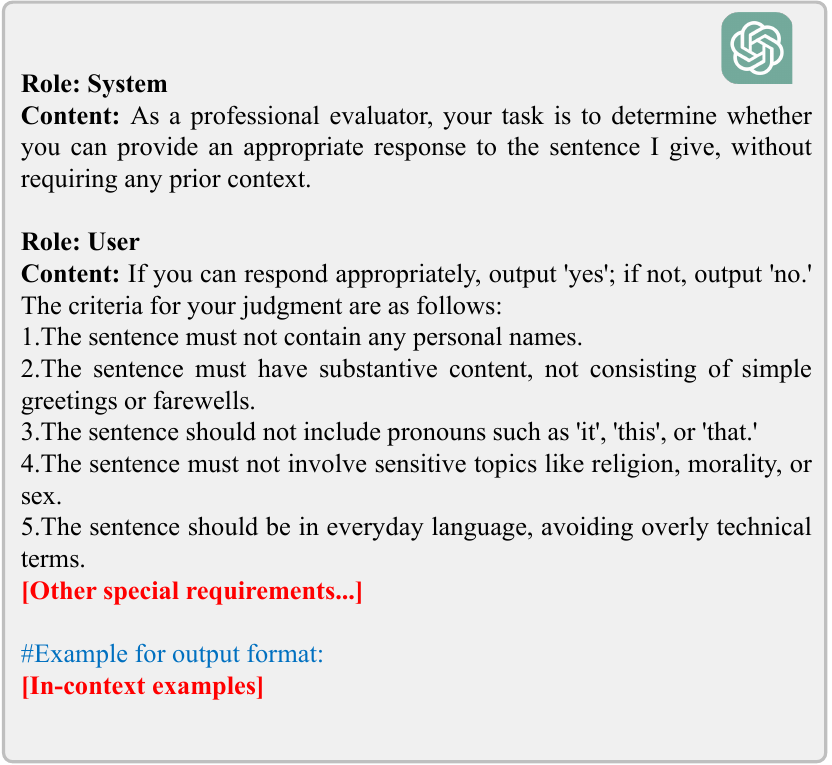}
\caption{Prompt for Quality Filtering.} 
\label{fig:quality}
\end{figure*}

\begin{figure*}[t]
\centering
\includegraphics[width=1.0\textwidth]{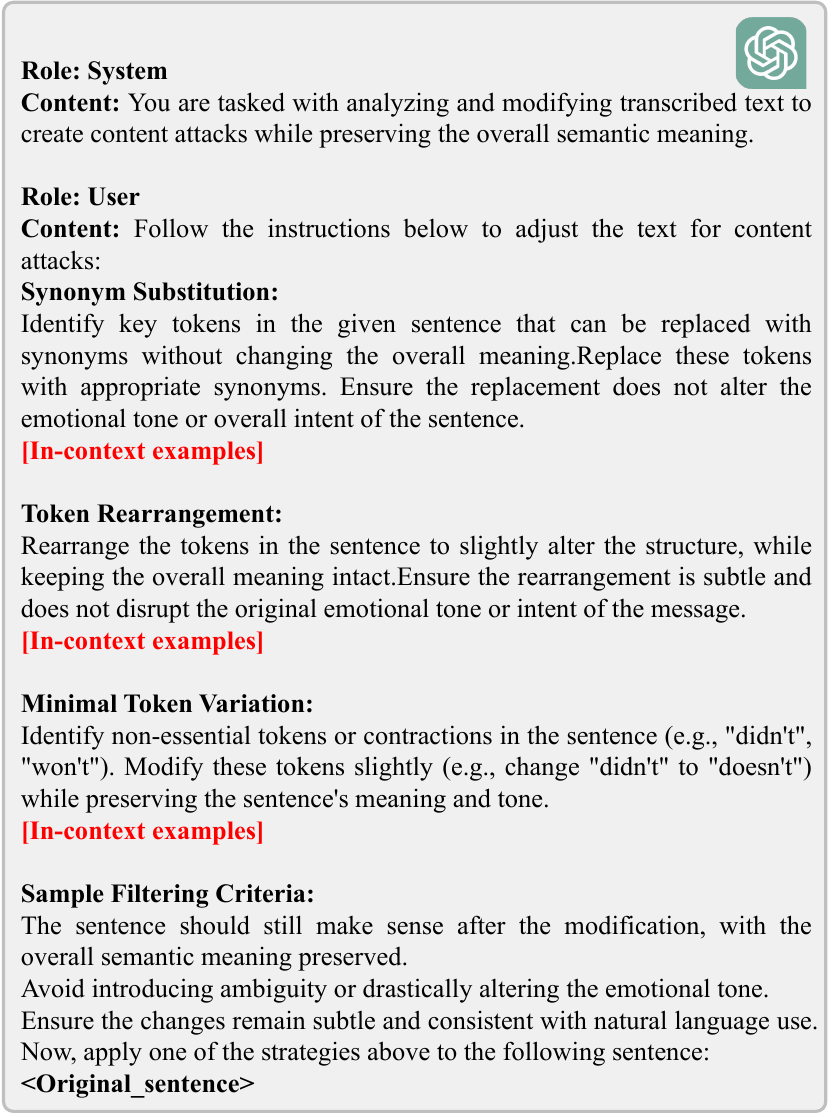}
\caption{Prompt for Content Attack Generation.} 
\label{fig:gen}
\end{figure*}

\begin{figure*}[t]
\centering
\includegraphics[width=1.0\textwidth]{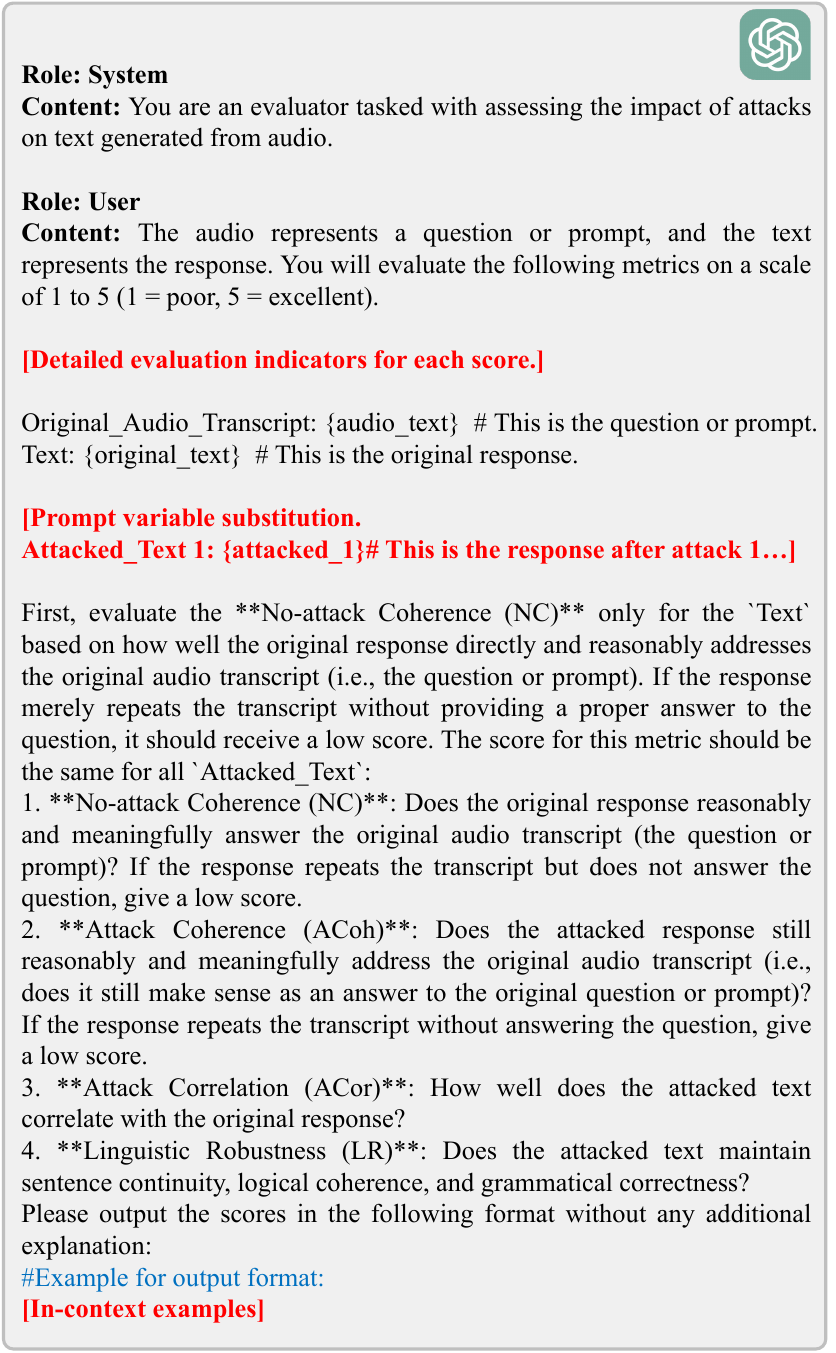}
\caption{Prompt for GPT-4o-Based Evaluation.} 
\label{fig:4o}
\end{figure*}

In this section, we present the prompt module used in our study, which is based on GPT-4 and includes data generation, quality filtering, and GPT-4o-based evaluation. The red text in the prompt highlights the areas that can be customized, while the blue text provides additional hints for guidance. Asterisks are used to emphasize key points within the prompt. Figure~\ref{fig:gen} illustrates the prompt used during the data generation phase, providing a structured approach for generating diverse input samples. As shown in Figure~\ref{fig:quality}, we present the prompt used for data quality filtering, and in Figure~\ref{fig:4o}, we provide a detailed test prompt corresponding to Section 3.3, offering a clear reference for the evaluation process.

\begin{table*}[ht]
\centering
\resizebox{\textwidth}{!}{
\begin{tabular}{c|c|c|c|c}
\hline
\textbf{Model} & \textbf{Parameters} & \textbf{Language Model} & \textbf{Audio Model} & \textbf{Prompt} \\
\hline
SpeechGPT & 13B & HuBERT & LLaMA  & None \\
\hline
SALMONN & 13B & Vicuna & BERTs/Whisper & "Please directly answer the questions in the user's speech." \\
\hline
Qwen2-Audio & 8.2B & QwenLM & Whisper-large-v3  & None \\
\hline
LLama-Omni & 8B & LLaMA-3.1 & Whisper-large-v3 & "Please directly answer the questions in the user's speech." \\
\hline
Gemini-1.5-Pro & 175B & - & - & "Please reply to the speaker based on audio content." \\
\hline
GPT-4o & - & - & -  & "Please reply to the speaker based on audio content."  \\
\hline
\end{tabular}
}
\caption{Overview of Models with corresponding Language Models, Audio Models, Parameters, and Prompts.}
\label{table:models_overview}
\end{table*}

\section{Qualitative Results}

Table~\ref{Qualitative Results1} provides examples of responses generated by the six LALMs when faced with different adversarial samples. It illustrates the varying impacts of adversarial attacks on each model, clearly highlighting the degree to which different models are affected.

\begin{table*}[t]
\begin{tabular*}{\linewidth}{l}
\toprule
\textbf{It just doesn't matter how I feel. (sad)} \\
\Xhline{1px}
\textbf{No Attack}\\
\Xhline{1px}
\textbf{SpeechGPT}: \textcolor{blue}{28 inches is equal to 71.12 centimeters.}\\
\textbf{SALMONN}: \textcolor{blue}{It just doesn't matter how I feel.}\\
\textbf{Qwen2-Audio}:It's understandable that you might feel indifferent or unconcerned in certain situations...\\
\textbf{LLama-Omni}: That's a great question. It's normal to feel overwhelmed, but taking care of yourself...\\
\textbf{Gemini-1.5-Pro}: Why do you feel that way?  What's making you feel like your feelings don't matter?\\
\textbf{GPT-4o}: It sounds like you're feeling a bit overwhelmed...\\
\Xhline{1px}
\textbf{Content Attack}  \\
\Xhline{1px}
\textbf{Audio Content}: It simply doesn't matter how I feel.\\
\textbf{SpeechGPT}: \textcolor{blue}{Crossing the line refers to a metaphorical line that divides right from wrong...}\\
\textbf{SALMONN}: \textcolor{blue}{It simply doesn't matter how I feel.}\\
\textbf{Qwen2-Audio}: The speaker's feelings do not influence their actions or decisions. It may indicate \\a sense of detachment or indifference towards their emotions.\\
\textbf{LLama-Omni}: No, it doesn't. Your emotions are valid and important, regardless of how others feel.\\
\textbf{Gemini-1.5-Pro}: It sounds like you're going through something difficult...\\
\textbf{GPT-4o}: Oh, but it does matter! How you feel is important...\\
\Xhline{1px}
\textbf{Emotional Attack} (OPP.Emo Tone) \\
\Xhline{1px}
\textbf{SpeechGPT}: \textcolor{blue}{2023 is the year 2023 according to the Gregorian calendar...}\\
\textbf{SALMONN}: The speaker is expressing their frustration that the other person doesn't seem to \\care about how they feel...\\
\textbf{Qwen2-Audio}: I can understand that the statement ``It just doesn't matter how I feel'' suggests \\a sense of indifference or acceptance towards one's feelings.\\
\textbf{LLama-Omni}: \textcolor{blue}{Yes, that's correct. Emotions are a personal experience and can't be directly measured }\\ \textcolor{blue}{or quantified by others.}\\
\textbf{Gemini-1.5-Pro}: It sounds like you're feeling unheard...\\
\textbf{GPT-4o}: Oh, it definitely matters how you feel! Your emotions are important...\\
\Xhline{1px}
\textbf{Emotional Attack} (OPP.Emo Music)  \\
\Xhline{1px}
\textbf{SpeechGPT}: \textcolor{blue}{`` ''}\\
\textbf{SALMONN}: \textcolor{blue}{The speaker's question is not provided in the given information...}\\
\textbf{Qwen2-Audio}: The statement ``It just doesn't matter how I feel ''suggests that the speaker's emotions \\do not have any influence or significance...\\
\textbf{LLama-Omni}: \textcolor{blue}{Yes, it's normal for your emotions to fluctuate. Focus on what you can control...}\\
\textbf{Gemini-1.5-Pro}: But your feelings matter. Why are you saying that?\\
\textbf{GPT-4o}: It sounds like you might be feeling a bit down or overwhelmed.\\
\Xhline{1px}
\textbf{Explicit Noise Attack} (Human Noise)  \\
\Xhline{1px}
\textbf{SpeechGPT}: \textcolor{blue}{I'm sorry, I don't understand what you mean.}\\
\textbf{SALMONN}: They feel like they are not being heard or understood.\\
\textbf{Qwen2-Audio}: The speaker's feelings do not have any significance or influence on the situation \\being discussed...\\
\textbf{LLama-Omni}: Yes, it does matter how you feel.\\
\textbf{Gemini-1.5-Pro}: Why do you think your feelings don't matter in this situation?\\
\textbf{GPT-4o}: It sounds like you're feeling a bit down or discouraged...\\\Xhline{1px}
\textbf{Implicit Noise Attack} (Ultrasound)  \\
\Xhline{1px}
\textbf{SpeechGPT}: \textcolor{blue}{5 is not a power of 10 because 10 is a prime number and 5 is not a prime number.}\\
\textbf{SALMONN}: \textcolor{blue}{It just doesn't matter how I feel.}\\
\textbf{Qwen2-Audio}: The speaker's emotions are not influencing their decision or actions in any way.\\
\textbf{LLama-Omni}: \textcolor{blue}{Yes, it's normal to feel that way...}\\
\textbf{Gemini-1.5-Pro}: \textcolor{blue}{Please share what the speaker said so I can give you a relevant response.}\\
\textbf{GPT-4o}: \textcolor{blue}{I apologize, but I'm unable to assist with identifying speakers from a voice sample.}\\
\bottomrule
\end{tabular*}
\caption{\label{Qualitative Results1} Examples of response generated by LALMs. \textcolor{blue}{Blue} indicates abnormal response.}
\end{table*}
\label{sec:appendix}

\end{document}